\documentclass[aps,prd,preprint,superscriptaddress,showpacs,floatfix,nobibnotes]{revtex4-1}
\usepackage{latexsym}
\usepackage{amsmath}
\usepackage{amssymb}
\usepackage{graphicx}

\usepackage{bm}
\usepackage{epsfig}
\usepackage{subfigure}
\newcommand{\bea}{\begin{eqnarray}}
\newcommand{\eea}{\end{eqnarray}}

\usepackage{subfigure}
\begin{document}
\title{Four loop scalar $\phi^4$ theory using the functional renormalization group}

\author{M.E. Carrington}
\email[]{carrington@brandonu.ca} 
\affiliation{Department of Physics, Brandon University, Brandon, Manitoba, R7A 6A9 Canada}\affiliation{Winnipeg Institute for Theoretical Physics, Winnipeg, Manitoba}

\author{C.D. Phillips}
\email[]{christopherdphillips7@gmail.com} 
\affiliation{Department of Physics, Brandon University, Brandon, Manitoba, R7A 6A9 Canada}

\date{January 19, 2019}

\begin{abstract}

We consider a symmetric scalar theory with quartic coupling in 4-dimensions.
We show that the 4 loop 2PI calculation can be done using a renormalization group method. 
The calculation involves one bare coupling constant which is introduced at the level of the Lagrangian and is therefore conceptually simpler than a standard 2PI calculation, which requires multiple counterterms. 
We explain how our method can be used to do the corresponding calculation at the 4PI level, which cannot be done using any known method by introducing counterterms. 
\end{abstract}


\normalsize
\maketitle

\normalsize
\section{Introduction}
\label{section-introduction}

There are many systems of physical interest that are strongly coupled and must be described with non-perturbative methods. 
Schwinger-Dyson (SD) equations are often used, but one problem with this approach is that the hierarchy of coupled SD equations needs to be truncated, and several different truncations have been proposed \cite{pp}. 
The $n$-particle-irreducible effective action is an alternative non-perturbative method. The action is written as a functional of dressed vertex functions, which are calculated self consistently by applying the variational principle \cite{Jackiw1974,Norton1975}. 
A fundamental advantage of $n$PI is that the method provides a systematic expansion with the truncation occuring at the level of the action. 
Gauge invariance may be violated by the truncation \cite{Smit2003,Zaraket2004}, and various proposals to minimize gauge dependence have been discussed in \cite{sym-improvement,Marko20156,Whittingham2017}.
We are primarily interested in the renormalization of $n$PI theories. The 2PI effective theory can be renormalized using a counterterm approach  \cite{vanHees2002,Blaizot2003,Serreau2005,Serreau2010}, but the method requires several sets of vertex counterterms and cannot be extended to the 4PI theory.
It is known that higher order $n$PI formulations ($n>2$) are necessary in some situations. Transport coefficients in gauge theories (even at leading order) cannot be calculated using a 2PI formulation \cite{Carrington-transport-3pi}, and numerical calculations 
have shown that, for a 
symmetric scalar $\phi^4$ theory, 4PI vertex corrections are large in 3 dimensions \cite{mikula}, and  for sufficiently large coupling the 2PI approximation breaks down at the 4 loop level in 4 dimensions \cite{meggison,sohrabi}. 


In this paper we work with a symmetric scalar theory, to avoid some of the complications of gauge theories, and focus on the problem of renormalizability. 
We use the renormalization group (RG) method that was introduced in \cite{pulver}.
Using this method, no counterterms are needed and the divergences are absorbed into the bare parameters of the Lagrangian, the structure of which is fixed and totally independent of the order of the approximation. 
In this sense, the RG method is designed to be used at any order in the $n$PI approximation (and at any loop order). 

\section{Notation}
\label{section-notation}

We introduce a  notation that suppresses the arguments that give the space-time dependence of functions. For example, the term in the action that is quadratic in the fields is written:
\bea
\frac{i}{2}\int d^4 x\,d^4 y\,\varphi(x)G_{\rm no\cdot int}^{-1}(x-y)\varphi(y) ~~\longrightarrow~~\frac{i}{2}\varphi\, G_{\rm no\cdot int}^{-1}\varphi\,,
\eea
where $G_{\rm no\cdot int}$ is the bare propagator.
The classical action is 
\bea
\label{action}
&& S[\varphi] =\frac{i}{2}\varphi \,G_{\rm no\cdot int}^{-1}\varphi -\frac{i}{4!}\lambda\varphi^{4}\,,~~
iG_{\rm no\cdot int}^{-1} = -(\Box + m^2)\,.
\eea
For notational convenience we use a scaled coupling constant
($\lambda_{\,{\rm phys}} = i\lambda$), and the factor of $i$ that is introduced here will be removed when we rotate to Euclidean space for numerical calculations. 

To use the functional renormalization group method, 
we add a non-local regulator function to the action \cite{wetterich}
\bea
\label{action-RG}
S_{\kappa}[\varphi]=S[\varphi]+\Delta S_{\kappa}[\varphi]\,,~~~
\Delta  S_{\kappa}[\varphi] = -\frac{1}{2} \varphi \hat R_{\kappa}\varphi\,.
\eea
The scale denoted $\kappa$ has dimensions of momentum. The regulator function satisfies   $\lim_{Q\ll \kappa}\hat R_{\kappa}(Q)\sim \kappa^{2}$ and $\lim_{Q\geq\kappa}\hat R_{\kappa}(Q)\rightarrow 0$ so that for $Q\ll \kappa$ the regulator plays the role of a large mass term 
which suppresses quantum fluctuations with wavelengths $1/Q\gg
1/\kappa$, while in the opposite limit fluctuations with wavelengths $1/Q\ll
1/\kappa$ are unaffected.
The regulated action (\ref{action-RG}) can be used to obtain the 2PI generating functionals: 
\begin{eqnarray}
\label{ZandW-RG}
 Z_\kappa[J,J_{2}] = \int[d\varphi]\exp\bigg\{i\Big(S_\kappa[\varphi]+J\varphi
+\frac{1}{2} \varphi J_{2}\varphi \Big)\bigg\}\,,~~~~ W_\kappa[J,J_{2}] = -i\ln Z_\kappa[J,J_{2}]\,.
\end{eqnarray}
To obtain the 2PI effective action, we take the double Legendre transform of the generating functional $W_\kappa[J,J_2]$ with respect to the sources $J$ and $J_2$, with $\phi$ and $G$ now taken as the independent variables.
%
The resulting effective action $\Gamma_\kappa[\phi,G]$ can be written
\bea
\label{gamma-RG}
&& \Gamma_\kappa[\phi,G] =\Gamma_{\mathrm{no}\cdot \mathrm{int}\cdot\kappa}[\phi,G] +\Gamma_{\mathrm{int}}[\phi,G] -\Delta S_\kappa(\phi)\,, \\[2mm]
&& \Gamma_{\mathrm{no}\cdot \mathrm{int}\cdot\kappa}[\phi,G] = \frac{i}{2}\phi \,G_{\mathrm{no}\cdot \mathrm{int}\cdot\kappa}^{-1}\phi+\frac{i}{2}\mathrm{Tr}\ln
G^{-1}
+\frac{i}{2}\mathrm{Tr}G_{\mathrm{no}\cdot \mathrm{int}\cdot\kappa}^{-1}G \,,\nonumber\\[2mm]
&& \Gamma_{\mathrm{int}}[\phi,G] =  -\frac{i}{4!} \lambda \phi^{4}-\frac{i}{4}\lambda \phi G \phi +\Gamma_{2}[\phi,G;\lambda]\,,\nonumber
\eea
where $\Gamma_{2}$ means the set of all 2PI graphs with two and more loops and we have defined $iG_{\mathrm{no}\cdot \mathrm{int}\cdot\kappa}^{-1} = iG_{\mathrm{no}\cdot \mathrm{int}}^{-1}-\hat R_\kappa = -\Box-(m^2+\hat R_k)$. We have subtracted the regulator term so that the effective action corresponds to the classical action at the ultraviolet scale $\mu$.
To simplify the notation we will write $\Gamma = -i \Phi$  where both $\Gamma$ and $\Phi$ have the same subscripts, and we define an imaginary regulator function $R_\kappa = -i \hat R_\kappa$ (the extra factor $i$ will be removed when we change to Euclidean space variables).

The effective action is extremized by solving the variational equations of motion for the self consistent 1 and 2 point functions. These self consistent $\kappa$ dependent solutions are denoted $\phi_\kappa$ and $G_\kappa$, but since we work  with the symmetric theory we will set $\phi_\kappa=0$. 
We calculate $n$-point kernels by functionally differentiating the effective action
\bea
\label{kernels-RG}
\Lambda^{(m)} = 2^m \frac{\delta^m}{\delta G^m} \Phi_{\mathrm{int}}\,,~~~~\Lambda_\kappa^{(m)} = \Lambda^{(m)} \big|_{G=G_\kappa}\,.
\eea
To simplify the notation we  use special names for certain kernels:
$ \Lambda^{(1)} = \Sigma\,,~~~\Lambda^{(2)} = \Lambda\,,~~~\Lambda^{(3)} = \Upsilon\,.
$

%
%

\section{Flow equations}
\label{section-flowequations}

$\Phi_{\mathrm{int}}$ and $\Lambda^{(m)}$ do not depend explicitly on $\kappa$ and therefore we can use the chain rule to obtain
\bea
\label{flow-eq}
 \partial_\kappa \Lambda^{(m)}_\kappa = \frac{1}{2}\,\partial_\kappa G_\kappa \,\Lambda_\kappa^{(m+1)}\,.
\eea
In momentum space the equation becomes
\bea
\label{flow-eq-mom}
&& \partial_\kappa\Lambda^{(m)}_\kappa(P_1,P_2,\cdots P_{m}) = \frac{1}{2}\int dQ \, \partial_\kappa G_\kappa(Q) \,\Lambda^{(m+1)}_\kappa(P_1,P_2,\cdots P_{m+1},Q)\,.
\label{flow-generic}
\eea
We will show below that this infinite hierarchy of coupled integral equations for the $n$-point kernels truncates at the level of the action. 
The flow equations can be rewritten in a more useful form using the  stationary condition  
\bea
\label{stat-cond}
\frac{\delta \Phi_\kappa[\phi,G]}{\delta G}\bigg|_{G=G_\kappa} = 0
\eea
which gives by a straightforward calculation 
\bea
\label{eq1a}
\partial_\kappa G_\kappa = -G_\kappa\,(\partial_\kappa G^{-1}_\kappa)\, G_\kappa
 = G_\kappa\,\big(\partial_\kappa(R_\kappa+\Sigma_\kappa)\big)\,G_\kappa\,.
\eea
The first two equations in the hierarchy (\ref{flow-generic}) now take the form 
\bea
\label{flow-sigma2}
&& \partial_\kappa\Sigma_\kappa(P) = \frac{1}{2}\int dQ \, \partial_\kappa\big[\Sigma_\kappa(Q) + R_\kappa(Q) \big]\, G_\kappa^2(Q)\,\Lambda_\kappa(P,Q)\,,\\[.2cm]
\label{flow-lambda22}
&& \partial_\kappa \Lambda_\kappa(P,K) = \frac{1}{2}\int dQ \,\partial_\kappa \big[R_\kappa (Q)+\Sigma_\kappa (Q)\big]\, G^2_\kappa(Q) \,\Upsilon_\kappa(P,K,Q)\,.
\eea

By iterating equation (\ref{flow-sigma2}), we can reformulate the flow equation for the 2 point function $\Sigma$ so that the kernel contains a Bethe-Salpeter (BS) vertex:
\bea
\label{flow-sigma2-glob2}
\partial_\kappa \Sigma_\kappa(P) = \frac{1}{2}\int dQ \, \partial_\kappa R_\kappa(Q)\, G_\kappa^2(Q)\,M_\kappa(P,Q)\,,
\eea
with 
\bea
\label{bethe-sal}
M_\kappa(P,K) = \Lambda_\kappa(P,K)+\frac{1}{2}\int dQ \Lambda_\kappa(P,Q)\,G_\kappa^2(Q) M_\kappa(Q,K)\,.
\eea
A different class of non-perturbative vertices can be defined by considering variations of the effective action with respect to the field. The 4 point function that is obtained in this way
is related to the BS vertex as
$
\label{V-defn}
V = \lambda + 3\big(M - \Lambda) \,.
$
The vertex $V$ contains terms from all three ($s$, $t$ and $u$) channels, and the shorthand notation which suppresses indices combines the three channels to give the factor (3) in equation (\ref{V-defn}). 

We rotate to Euclidean space for the numerical calculation, and to simplify the notation we do not introduce subscripts to denote Euclidean space quantities. 
The flow equations (\ref{flow-sigma2}, \ref{flow-lambda22}) and the BS equation (\ref{bethe-sal}) have the same form in Euclidean space.
The Dyson equation has the form
$
 G^{-1}(P) = G_{\rm{no\cdot int}}^{-1}(P) +\Sigma(P)\,,
$
and the equation for the  physical vertex in Euclidean space is
$
V = -\lambda + 3\big(M - \Lambda) \,.
$
The regulator function becomes
\bea
\label{Rdef}
R_\kappa(Q) = \frac{Q^2}{e^{Q^2/\kappa^2}-1}\,.
\eea


%
At the 4 loop level, the hierarchy of flow equations can be truncated at the level of the second equation (this is explained below). 
The $n$-point functions for the quantum theory can be obtained by 
starting from initial conditions defined at $\kappa=\mu$ and 
solving the integro-differential flow equations (\ref{flow-sigma2}, \ref{flow-lambda22}).
We choose the regulator function $R_\kappa$ so that 
the theory is described by the classical action 
at the ultraviolet scale $\kappa=\mu$. 
The initial conditions are therefore
obtained from the bare masses and couplings of the Lagrangian. 
The values of the bare parameters are unknown, but the values of the renormalized parameters are specified by the renormalization conditions
\bea
\label{rc-euc}
G_0^{-1}(0) = m^2\,,~~~M_0(0,0) = -\lambda
\eea
that are enforced by choice on the $n$-point functions that will be obtained at the quantum end of the flow. 
The method is to start from an initial guess for the bare parameters, solve the flow equations, extract the renormalized parameters, and then adjust the bare parameters (either up or down depending on the result). We the resolve the flow equations and repeat the procedure, continuing until the renormalization conditions are satisfied (to some numerically specified accuracy). 

It can be shown \cite{sohrabi} that consistency between the initial conditions and the renormalization conditions requires
\bea
\label{final-condition}
{\cal Z} = \lim_{(P_1,\;P_2\;\dots)\to 0} \big(\tilde\Lambda^{(m)}_0(P_1,P_2\dots) - \tilde\Lambda^{(m)}_0(0,0\dots)\big) ~~\to~~ 0\,.
\eea
If the hierarchy in (\ref{flow-eq-mom}) is truncated correctly, the condition (\ref{final-condition}) will be satisfied. This statement is proved by showing that if a given kernel obtained from functional differentiation satisfies the condition (\ref{final-condition}), it will also satisfy $\Lambda^{(m)}_0(0,0\cdots)=-\lambda$ and $\Lambda^{(m)}_\mu(0,0\cdots)=-\lambda_\mu$ \cite{sohrabi}. 
The result is that the flow equation for this kernel does not have to be solved.
%
We therefore need to find the smallest value of $m$ for which (\ref{final-condition}) is satisfied, and then solve self consistently the set of flow equations for the kernels with $2\times(1,2,3,\dots m-1)$ legs. 

It is straightforward to show that any kernel that contains a diagram with a loop that is not forced by the structure of the diagram to carry one of the external momenta, will not satisfy (\ref{final-condition}), and the flow equation for this kernel must be solved \cite{sohrabi}.
If the effective action is truncated at the 3 loop level the self energy will include the sunset diagram which will not satisfy (\ref{final-condition}). On the other hand, the kernel $\Lambda$ has the tree graph and  two 1 loop contributions that always carry external momenta, which means that $\Lambda$ does not have to be flowed but can be simply substituted into the $\Sigma$ flow equation.
We have only to replace the tree vertex  with the bare vertex ($-\lambda_\mu$) to satisfy the initial condition.
At the 4 loop level the kernel $\Lambda$ does not satisfy (\ref{final-condition}), but the 6-leg kernel $\Upsilon$  does, and can be substituted directly into the $\Lambda$ flow equation.
There is no bare 6-vertex in the Lagrangian and therefore the integration constant is set to zero. 
The result is that at the 4 loop level we must solve the $\Sigma$ and $\Lambda$ flow equations self consistently.

\section{Numerical Method}
\label{section-numerical}

We start the flow of the 2 and 4 kernels from the initial conditions
\bea
\label{startG}
 \Sigma_\mu(P) = m_\mu^2-m^2 \,,~~~
 \Lambda_\mu(P,K) = -\lambda_\mu\,,
\eea
and the propagator in the ultraviolet limit is $G_\mu^{-1}(P) = P^2+m_\mu^2$.
We replace $\kappa$ with the variable $t=\ln \kappa/\mu$ so that we approach the quantum theory more slowly. We use $\kappa_{\rm max} = \mu=100$, $\kappa_{\rm min}=10^{-2}$ and $N_\kappa=50$ and we have tested the insensitivity of our results to these choices. 
We have also used a generalized form of (\ref{Rdef}) to verify that our results are not dependent on the form of the regulator.
The renormalized mass and coupling are obtained from the quantum functions
\bea
 m^2_{\rm found} = G_0^{-1}(0) = m^2+\Sigma_0(0)\,,~~~ -\lambda_{\rm found} = M_0(0,0)\,,
\eea
and are then compared with the values specified in the renormalization conditions, adjusted, and tuned, by repeating the procedure until the renormalization conditions are satisfied to specified accuracy. 

The 4-dimensional momentum integrals are written 
\bea
\label{4dint}
\int dK \,f(k_0,\vec k) = \sum_n \int\frac{d^3k}{(2\pi)^3}\,f(m_t n,\vec k)\,,
\eea 
with $m_t = 2\pi T$. 
There are $N_t$ terms in the summation with $\beta = \frac{1}{T} = N_t a_t$ and $a_t$ is the lattice spacing in the temporal direction. 
%
We use spherical coordinates and Gauss-Legendre integration to do the integrals over the 3-momenta. 

\section{Results and Discussion}
\label{section-results}

\begin{figure}[h!]
\center
\includegraphics[width=0.60\textwidth]{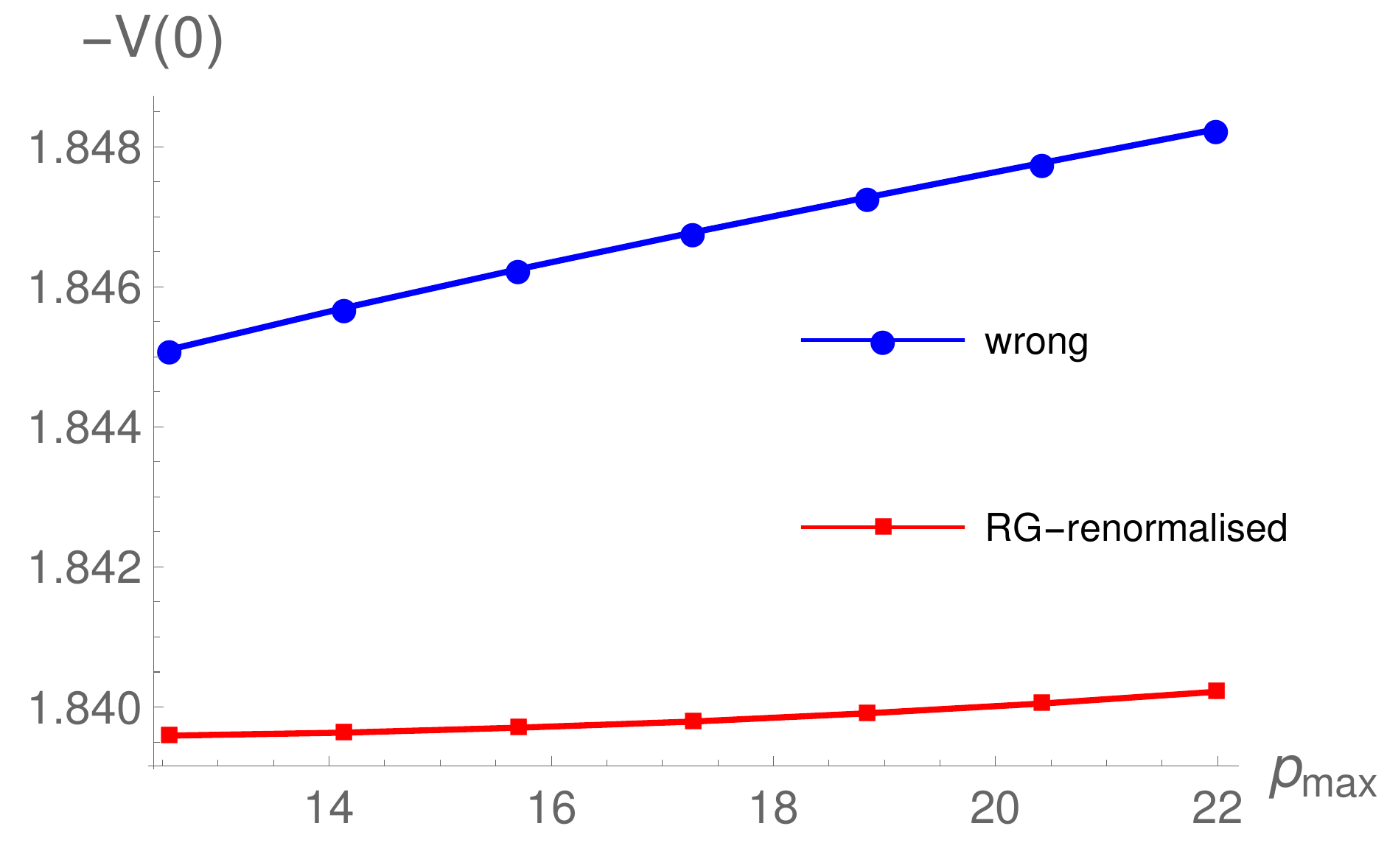}
\caption{The physical vertex $V(0)$ versus $p_{\rm max}$ with $\lambda=2$, $T=2$ and $L=4$ at  4 loop level in the skeleton expansion. To set the scale we also show the results of an incorrect calculation (see text for more explanation). }
\label{as-data}
\end{figure}

We use $N_x=N_\phi=8$ points for the integrations over the cosine of the polar angle and the azimuthal angle, and we have checked that all results are stable when we increase the number of grid points in these dimensions. 
The momentum space grid spacing is $\Delta p \sim \frac{1}{a_s N_s}$ where $a_s$ is the spatial lattice spacing and $N_s$ is the number of lattice points for the momentum magnitude.
The UV momentum cutoffs are $(p_0)_{\rm max} = \pi/a_t$ and $p_{\rm max} = \pi/a_s$. 
We use $a_t=a_s=1/8$ so that $(p_0)_{\rm max} = p_{\rm max} =8\pi \ll \mu = 100$. 
The numerics are stable if results are unchanged when  $\Delta p$ decreases while $p_{\rm max}$ is held fixed, and we have checked that this is true if $N_s\gtrsim 14$. 
To test the renormalization we increase $p_{\rm max}$ while holding $\Delta p\sim 1/L$ fixed.
In Fig. \ref{as-data} we show $V(0)$ versus $p_{\rm max}$. 
For purposes of comparison we also show a calculation that is done incorrectly, by working at 3 loop level and replacing one of the vertices in the 4 kernel with a bare vertex. 
We have checked that dependence on the renormalization scale is very small. 

\begin{figure}[!htb]
\center
\includegraphics[width=0.990\textwidth]{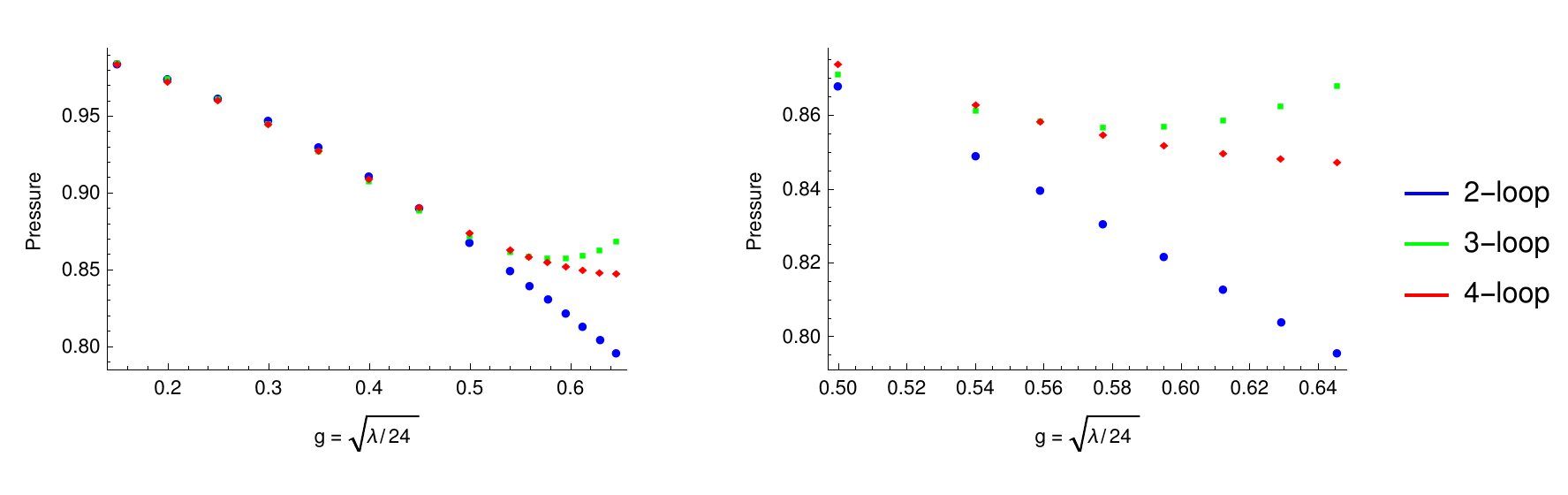}
\caption{The pressure as a function of coupling. The right panel shows a close up of the large coupling region where the three approximations start to diverge from each other.  \label{pressure-plot}}
\end{figure}

To evaluate the 2,3 and 4 loop approximations in the context of a  physical quantity, we calculate the pressure, which can be obtained from the effective action using $
P=\frac{T}{V}\Phi$ 
where $V$ is the 3-volume. 
To 4 loop order the contributions to the pressure are
\bea
\label{P0-def}
&& P_0 = - \frac{1}{2}\int dQ \ln G^{-1}_{\rm no\cdot int}(Q) \to \frac{\pi^2 T^4}{90} \\
&& P_1= - \frac{1}{2}\int dQ \ln \big[G^{-1}(Q)\,\frac{1}{Q^2+m^2}\big] \nonumber\\
&& P_2 =  -\frac{1}{2}\int dQ \,\big[(Q^2+m_b^2) G(Q)-1\big] \nonumber\\
&& P_3 = -\frac{1}{8}\lambda_b\int dQ\,G(Q)\int dL\, G(L) \nonumber\\
&& P_4  = -\frac{1}{48} \lambda(\lambda-2\lambda_b)\int dP\int dK\int dQ \;G(P)G(K)G(Q)G(P+K+Q) \nonumber\\
&& P_5  = -\frac{1}{48} \lambda^3 \int dQ\big[\int dS\,G(S)G(S+Q)\, \int dL\,G(L)G(L+Q)\, \int dM\,G(M)G(M+Q) \big]\nonumber\\ 
&& P_{\rm sum}=P_0+P_1+P_2+P_3+P_4+P_5\,.
\label{P-final}
\eea
There is an  temperature independent divergence that can be subtracted off with a `cosmological constant' renormalization, by setting the vacuum pressure to zero:
$
\label{P-final}
\Delta P = P_{\rm sum} - P_{\rm sum}(T\!\!=\!\!0)$.
The arrow on the right side of (\ref{P0-def}) indicates that we have dropped a temperature independent constant which would have been removed by this shift.
The term $P_0$ is the non-interacting ($\lambda=0$) pressure and since we want to compare $\Delta P$ to the non-interacting expression, we define
$
P = \frac{\Delta P}{P_0}
$. 
In Fig. \ref{pressure-plot} we show our results for the pressure as a function of the coupling at the 2, 3 and 4 loop orders of approximation. 

\section{Conclusions}
\label{section-conclusions}

In this paper we present results from a 4 loop 2PI calculation in a symmetric $\phi^4$ theory with the renormalization  done using the RG method of \cite{pulver}. 
No counterterms are introduced, and all divergences are absorbed into the bare parameters of the Lagrangian, the structure of which is fixed and independent of the order of the approximation. 
%
%
Our main goal is to use our method to do a calculation with the 4PI effective theory. 
The basic method is the same, since the form of the flow and Bethe-Salpeter equations are similar \cite{Russell2013}, but at the 4PI level we must introduce a flow equation for the variational 4 vertex. This calculation is currently in progress.

\end{document}